\documentclass[twocolumn,preprintnumbers,amsmath,amssymb]{revtex4}

\usepackage{graphicx}
\usepackage[latin1]{inputenc}
\usepackage{amsmath}
\usepackage{amsfonts}
\usepackage{amssymb}
\usepackage{amsthm}
\usepackage{amsxtra}
\usepackage{fancyhdr}
\usepackage{mathrsfs}
\newcommand{\be}{\begin{equation}}
\newcommand{\ee}{\end{equation}}
\newcommand{\ben}{\begin{eqnarray}}
\newcommand{\een}{\end{eqnarray}}

\newcommand{\Nqi}{\bar{n}^*_i}

\begin{document}
\draft
\title{Thermodynamic Consistency of the  $q$-Deformed Fermi-Dirac 
Distribution in Nonextensive Thermostatics}

\author{J. M. Conroy$^1$, H.G. Miller$^1$ and  A.R. Plastino$^{2,\,3}$ }

\affiliation{$^1$Department of Physics, SUNY Fredonia, Fredonia, New York,
USA}

\affiliation{$^2$Instituto Carlos I de F\'{\i}sica Te\'orica y Computacional,
Universidad de Granada, Granada, Spain}

\affiliation{$^3$National University La Plata, CREG-UNLP-CONICET,
C.C. 727, 1900 La Plata, Argentina}

\date{\today}

\begin{abstract}

 The $q$-deformed statistics for fermions arising within the 
non-extensive thermostatistical formalism has been applied to
the study of various quantum many-body systems recently. The aim
of the present note is to point out some subtle difficulties 
presented by this approach in connection with the problem of 
thermodynamic consistency. Different possible ways to apply 
the $q$-deformed quantum distributions in a thermodynamically 
consistent way are considered.


  \end{abstract}  


\maketitle

 The $q$-deformed quantum distributions 
\cite{BD93,BDG95,PPP95,C96,C96b,TBD97,TBD98,U99,TT00,SAA10} 
inspired on the nonextensive thermostatistical 
formalism \cite{T09,TG04,TGS05} have been the focus 
of considerable attention in recent years (see 
\cite{RW09,CM08,PSA07,MKTPP06,TMT03,UMK01} and 
references therein). An interesting recent development 
along these lines was the formulation by Pereira,
Silva, and Alcaniz (PSA) of a $q$-deformed equation 
of state for relativistic nuclear matter within 
Walecka's phenomenological relativistic approach 
\cite{PSA07}. The PSA equation of state may be 
relevant for the study of nuclear matter in 
neutron stars.  Strictly speaking, however, the PSA
equation of state as derived in \cite{PSA07}
is not thermodynamically consistent. Here we will
consider possible solutions to this difficulty.
As we are going to demonstrate, the PSA approach 
can be implemented in a thermodynamically consistent 
way either by adopting a $q$-nonlinear form
for thermodynamical quantities like the total 
energy $E$ or the total number of particles $N$ 
that in the standard thermostatistical formalism
are linear functions of the (mean) occupation numbers 
or, alternatively, by recourse to a different choice 
of the entropy functional. This last procedure can,
in turn, be implemented in two different ways.
One possible re-definition of the entropy, accounting
for the relevant equations of constraint, yields 
the same quantum distribution functions as solutions 
of the entropic variational principle as the one
employed by PSA, and it preserves thermodynamic consistency.

Consider the $q$-deformed Fermi-Dirac distribution used
by PSA which has the form

\ben \label{qfernewc}
\bar{n}_i \,
 &=& \, \frac{1}{1+[1+(\tilde 
q-1)(\alpha+\beta\,\epsilon_i)]^{\frac{1}{\tilde q-1}}} \cr
&=& \frac {1} {1+[1+(\tilde q-1)\,
\beta(\epsilon_i-\mu)]^{\frac {1} {\tilde 
q-1}}},
\een

\noindent
where $\epsilon_i$ are the single particle energies, 
$\beta = 1/kT$ ($T$ being the absolute temperature 
and $k$ denoting Boltzmann's constant), 
$\mu=-\frac{\alpha}{\beta}$ is the 
chemical potential, and

\be \label{tildeq}
\tilde q \, = \, \left\{\begin{array}{ll}
q, &\quad {\rm if}\quad \alpha+\beta\,\epsilon_i > 0
\vspace{0.5cm} \\
2-q, &\quad {\rm if}\quad \alpha+\beta\,\epsilon_i \leq 0.
\end{array} \right.
\ee

\noindent
As shown in \cite{TPM05}, the Fermi mean occupation 
numbers (\ref{qfernewc}) can be obtained from a 
maximum entropy principle based on the entropic 
measure 

\begin{equation} \label{qentrop3}
S_q^{(F)}=\sum_i C_q(\bar{n}_i),
\end{equation}

\noindent
where the function $C_q(x)$ is defined by

\begin{equation} \label{cq}
C_q(x)=\left\{\begin{array}{ll}
\left(\frac {x-x^q} {q-1}\right)+\left(\frac{(1-x)-(1-x)^q} {q-1}\right) &\quad 
{\rm if}\quad x\leq \frac{1}{2}\vspace{0.5cm} \\
\left(\frac {x-x^{2-q}} {1-q}\right)+\left(\frac{(1-x)-(1-x)^{2-q}} 
{1-q}\right) &\quad {\rm if}\quad x>\frac{1}{2}
\end{array} \right.
\end{equation}

\noindent
The optimization of the entropic measure (\ref{qentrop3})
under appropriate constraints corresponding to the total 
number of particles $N$ and the total energy $E$ leads to 
the variational problem  

\be \label{variatiq}
\delta \!\! \left\{\!\frac{1}{k}S_q^{(F)}[\bar{n}]+\alpha\left(N\!-\!\sum_i 
\bar{n}_i^q\right) \!+ \!
\beta\!\left(E\!-\!\sum_i \epsilon_i\,\bar{n}_i^q\right)\!\right\} \!
 = \!0,
\ee

\noindent
where $\alpha$ and $\beta$ denote the 
Lagrange multipliers associated, respectively, 
with the aforementioned two constraints. 
The solution of equation (\ref{variatiq})
is given by the $q$-deformed Fermi distribution 
(\ref{qfernewc}). In the limit case $q\to 1$, the entropic
functional (\ref{qentrop3}) reduces to the well
known Fermi functional

\begin{equation} \label{standard1}
S = -\sum_i \Bigl[\bar{n}_i\ln \bar{n}_i +  
(1 - \bar{n}_i)\ln (1 -
\bar{n}_i)\Bigr],
\end{equation}

\noindent 
and the $q$-distribution (\ref{qfernewc})
reduces to the standard Fermi distribution.

\noindent
 Now, the most fundamental requirement of a thermostatistical
formalism is thermodynamical consistency. That is, the 
formalism must comply with the standard thermodynamical 
relationships among thermodynamical variables such as entropy, 
energy, temperature, etc. For instance, one requires
the well known relationship

\be \label{qutino}
\left(
\frac{\partial S}{\partial E} 
\right)_{V,N}
\, = \, \frac{1}{T},
\ee
 
\noindent
between the entropy, the energy, and the temperature
of a thermodynamical system at equilibrium to be
satisfied. It can be shown (see \cite{TPM05,PPMU04} and 
references therein for details) that any thermostatistical formalism
constructed on the basis of the constrained externalization of an 
entropic functional (that is, following Jayne's maximum entropy 
prescription) complies with the thermodynamical relationships (which, 
in the context of Jayne's' maxent formulation are usually referred 
to as Jaynes' relationships). To obtain a thermodynamically 
consistent formulation one has to make the appropriate 
identifications between relevant constraints and extensive 
thermodynamical quantities, on the one hand, and between the 
corresponding Lagrange multipliers and appropriate intensive 
thermodynamical quantities, on the other one. In the case of
the formalism based upon the entropic variational principle
(\ref{variatiq}) the appropriate identifications are

\ben \label{qit}
\sum_i \bar{n}_i^q &\rightarrow & N, \cr
\sum_i \bar {n}_i^q \,\epsilon_i &\rightarrow & E, 
\een

\noindent
and

\ben \label{qot}
\beta & \rightarrow & 1/kT, \cr
-\alpha/\beta & \rightarrow & \mu.
\een

\noindent
The functional (\ref{qentrop3}), of course, 
is to be identified with the entropy of the
system. It is plain from (\ref{qit})
that, in order to compute physical quantities 
in a thermodynamically consistent way, one
must not use directly the particle distribution
given by eq. (\ref{qfernewc}) (as is done
in \cite{PSA07}. See eqs.(17-20) in \cite{PSA07}) 
but, instead, use the effective 
particle distribution 

\be \label{qfeo}
\bar{n}_i^q \,
= \left(1+[1+(\tilde q-1)\,
\beta(\epsilon_i-\mu)]^{\frac{1}{\tilde q-1}}\right)^{-q},
\ee

\noindent
with $\tilde q$ defined as in (\ref{tildeq}). 
For instance, if the energy of an ideal Fermi 
gas is to comply with the basic thermodynamical 
relationship (\ref{qutino}), it has to be
computed (according to (\ref{qit})) as

\be
E \, = \, \sum_i
\epsilon_i \, 
\left(1+[1+(\tilde q-1)\,
\beta(\epsilon_i-\mu)]^{\frac{1}{\tilde q-1}}\right)^{-q}.
\ee

\noindent
It must be stressed that making the identification 
(\ref{qit}) does not imply any severe conceptual
difficulty, since the ${\bar n_i}$ are not probabilities and,
consequently, are not normalized to unity (see \cite{CHLW09}
for a similar situation arising in connection with
the $q$-generalization of the classical Boltzmann distribution). 
In fact, it is possible to reformulate the variational principle
(\ref{variatiq}) in terms of linear constraints, by recourse 
to an appropriate re-definition of the entropic functional.
Indeed, if one introduces the new entropy,

\begin{equation} \label{qentropnp}
{\tilde S}_q^{(F)}=\sum_i C_q(\bar{n}_i^{1/q}),
\end{equation}

\noindent
with the function $C_q$ still defined as in (\ref{cq}), then
the variational principle

\be \label{variatiqnp}
\delta \!\! \left\{\!\frac{1}{k} {\tilde S}_q^{(F)}[\bar{n}]+\alpha\left(N\!-\!\sum_i 
\bar{n}_i\right) \!+ \!
\beta\!\left(E\!-\!\sum_i \epsilon_i\,\bar{n}_i\right)\!\right\} \!
 = \!0,
\ee

\noindent
is equivalent to the variational
principle (\ref{variatiq}). The solution
to this variational problem is given by
the mean occupation numbers

\be \label{qlindo}
\bar{n}_i \,
= \left(1+[1+(\tilde q-1)\,
\beta(\epsilon_i-\mu)]^{\frac{1}{\tilde q-1}}\right)^{-q},
\ee

\noindent
which are given by the same expression as the one
given in (\ref{qfeo}).  The thermodynamical 
quantities $N$ and $E$ are now expressed in terms
of the mean occupation numbers (\ref{qlindo}) 
using the standard linear forms.


 There is another possible modification of the maxent variational
problem that can be implemented in order to recover
thermodynamic consistency. One can redefine the entropy 
functional to be

\begin{equation}\label{redefine}
 S_{q}^{*}=S_{q}+    
\alpha_1\!\sum_i 
(\bar{n}_i\!- \!\bar{n}_i^q) \!+ \!\beta_1\!\!\sum_i\epsilon_i\,
( \bar{n}_i\!-\!  \bar{n}_i^q).
\end{equation}

\noindent
In this case the variation of the entropy functional 
with the standard constraints $N=\sum_i \bar{n}_i$ 
and $E=\sum_i \epsilon_i \bar{n}_i$ is given by

\begin{equation}
 \delta\!\left\{\!S_{q}^{*}+\left(N\!-\alpha\!\sum_i 
 \!\bar{n}_i\right)\!+\left(E\!-\!\beta\!\!\sum_i 
 \epsilon_i\,\!  \bar{n}_i\!\right)\right\}\!=\!0\label{var2}
\end{equation}

\noindent
where $\alpha_1$ and $\beta_1$ are constants as q to be determined. 
Now if the following choices, $\alpha_1=\alpha$ and $\beta_1=\beta$ 
are made, the solution to the variational problem given in eq(\ref{var2}) 
is formally given by eq(\ref{qfernewc}) where $\alpha$ and $\beta$ 
are now determined by the standard equations of constraint since both 
variational equations, (\ref{variatiq}) and (\ref{var2}), are equivalent. 

However, in order for this to be a useful result, it is essential
that thermodynamic consistency is also preserved. 
In order to verify this it is necessary
to evaluate $(\frac{\partial S^*_q}{\partial E})_{VN}$.  
This can be done in the following manner,

\begin{eqnarray}
 \left(\frac{\partial S^*_q}{\partial E} \right)_{VN}&=&
 \left(\frac{\partial S^*_q}{\partial \Nqi} \right)_{VN} \Big/
\left(\frac{\partial E}{\partial \Nqi} \right)_{VN} \cr
&=&
\frac{q n_i^{q-1}}{\epsilon_i}
\left[ 
\beta \epsilon_i +\beta \epsilon_i
\left( \frac{1}{q n_i^{q-1}}-1 \right)
\right] \cr
&=&\beta
\end{eqnarray}

\noindent
as required.

Thermodynamic consistency is therefore preserved when 
the following identifications are made

\ben \label{qit1}
\sum_i \bar{n}_i &\rightarrow & N, \cr
\sum_i \bar {n}_i\,\epsilon_i &\rightarrow & E, 
\een

\noindent
and

\ben \label{qot1}
\beta & \rightarrow & 1/kT, \cr
-\alpha/\beta & \rightarrow & \mu.
\een

\noindent
It should also be noted that in the limit $q\to 1$, the last two 
terms in (\ref{redefine}) cancel and it again reduces to 
(\ref{standard1}). Requiring $S_{q}^{*} \to 0$ in the limit 
$T \to0$ places the restriction $q<2$.

\begin{figure}[hb]
             \includegraphics[width=3.2in,]{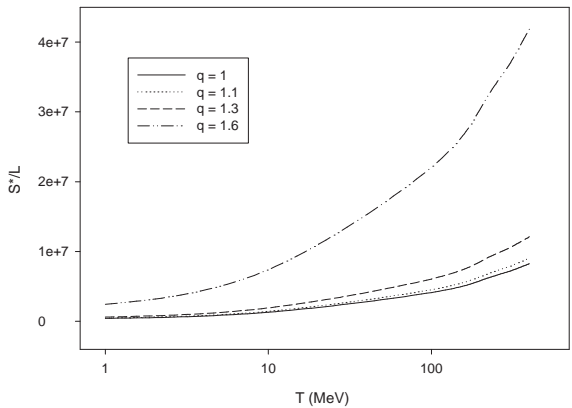}
             \caption{$S^{*}/L$ for m = 1 MeV and $\delta =10^{9}$MeV.}
          \label{fig1}
       \end{figure}

To illustrate the behavior of this redefined entropy functional, 
consider the toy model of a one-dimensional Fermi gas.  
In this case the entropy density (entropy per unit length) 
can be calculated from

\begin{equation}
 \frac{S^{*}}{L}= \int_{0}^{\delta}[C_{q}(\bar{n})+
 (\epsilon-\mu)\beta(\bar{n}-\bar{n}^{q})]g(\epsilon)d\epsilon
\end{equation}

\noindent
where $\delta$ is a high energy cutoff introduced for numerical 
purposes, and $g(\epsilon)d\epsilon= 
\frac{\sqrt{2m}}{\pi\hbar}\frac{1}{\sqrt{\epsilon}}d\epsilon$ 
is the one-dimensional density of states per unit length. 
For $m=1 {\rm MeV}$ and $\delta=10^9$ MeV, 
FIG.\ref{fig1} shows $S^{*}/L$ for various values of $q$.  
The entropy density increases at a given temperature for 
increasing values of $q$ but remains a strictly convex function.  
Thus, this redefinition introduces no additional structure 
into the entropy density. \\

In the present work we have considered the q-deformed quantum entropy functional  given in terms 
of   Fermi-Dirac  distribution functions.  Similar results can be obtained for quantum systems given in terms of Bose-Einstein distribution systems.   It also is interesting to note that analogous results can easily be obtained for  the q-deformed classical entropy functional. Modifying the entropy functional  admits the possibility of using  standard linear constraints such that thermodynamic consistency is preserved.

Summing up, our main conclusions are the following.
\\

\begin{itemize}
\item{First, a straightforward application of the variational
principle (\ref{variatiq})  is not consistent, from the 
thermodynamical point of view, with the use of the standard
identifications $N \rightarrow \sum_i \bar{n}_i$ and 
$E \rightarrow \sum_i \epsilon_i \bar{n}_i$ for the total
number of particles and the total energy, respectively
(a similar problem occurs with other thermodynamical
quantities that in the standard thermostatistical
formalism are expressed as linear functions of the 
mean occupation numbers).}

\item{Thermodynamic consistency can be recovered by using the
identifications $N \rightarrow \sum_i \bar{n}_i^q$ and 
$E \rightarrow \sum_i \epsilon_i \bar{n}_i^q$. This approach does
not lead to serious conceptual problems, because the mean occupation
numbers are not probabilities and are not normalized to one.}

\item{Alternatively, thermodynamic consistency can be recovered 
by appropriately redefining the entropy functional. In the present
work we have considered two alternative ways of implementing
this last procedure.}
\end{itemize}



Thermodynamical consistency is certainly a strong and fundamental
constraint in the development of extended or generalized 
thermostatistical formalisms of physical significance.
However, in the case of the $q$-deformed Fermi-Dirac statistics, 
this requirement alone does not determine unequivocally a unique
non-extensive generalization of the standard statistics for 
fermions. As we have pointed out, there are 
several ways to implement a thermostatistical formalisms for the 
$q$-deformed Fermi-Dirac distribution in a thermodynamically consistent 
way. Only when more experimental data is available, and more applications
to concrete quantum many-body systems are investigated,  will one 
definitively be able to ascertain which choices of the concomitant 
entropy functional and constraints should be used and under which 
circumstances. Any further developments along this lines will be
very welcome.

\acknowledgements
We would like to thank J. Cleymans for bringing this to our attention
This work was partially supported by 
the programs No. FQM-2445 and No. FQM-207
of the Junta de Andalucia (Spain).

\end{document}